\documentclass[a4paper,fleqn,usenatbib,useAMS,usefloat]{mnras}
\usepackage{mathptmx}

\usepackage[T1]{fontenc}
\usepackage{ae,aecompl}
\usepackage{graphicx}	
\usepackage{amsmath}	
\usepackage{amssymb}	
\usepackage{subfig}
\usepackage{ulem}
\usepackage{epstopdf}
\usepackage{physics}
\usepackage{makecell}

\newcommand\gaia{{\it Gaia}}
\newcommand\hipparcos{{\it Hipparcos}}

\newcommand{\qmin}{\mathcal{Q}_{y}}

\newcommand{\msun}{\textsc{m}_\odot} 
\newcommand{\mjup}{\textsc{m}_{\rm Jup}}
\newcommand{\MP}{$\mathcal{P}$-$\mathcal{M}_2$}

\def\msun{\ifmmode {\textsc M_{\odot}}    \else ${\textsc M_{\odot}}$ \fi}
\def\lsun{\ifmmode {\textsc L_{\odot}}    \else ${\textsc L_{\odot}}$ \fi}
\def\rsun{\ifmmode {\textsc R_{\odot}}    \else ${\textsc R_{\odot}}$ \fi}
\def\mjup{\ifmmode {\textsc M_{\textsc{Jup}}} \else ${\textsc M_{\textsc{Jup}}}$ \fi}
\def\rjup{\ifmmode {\textsc R_{\textsc{J}}} \else ${\textsc R_{\textsc{J}}}$ \fi}

\title[Boundaries of the Brown-Dwarf Desert as Seen with APOGEE]
{Study of the Mass-Ratio Distribution of Spectroscopic Binaries. II. The Boundaries of the Brown-Dwarf Desert as Seen with the  APOGEE Spectroscopic Binaries}
\author[Shahaf \& Mazeh]{
S. Shahaf\thanks{E-mail: sahar@wise.tau.ac.il}
and
T. Mazeh 
\\
School of Physics and Astronomy, Tel Aviv University, Tel Aviv 69978, Israel
}

\date{Accepted XXX. Received YYY; in original form ZZZ}

\begin{document}
\label{firstpage}
\pagerange{\pageref{firstpage}--\pageref{lastpage}}
\maketitle

\begin{abstract}
Analysis of APOGEE DR12 stellar radial-velocities by 
Troup et al.~(2016) 
affirmed the existence of the well-known Brown-Dwarf Desert (BDD). They detected a dearth of spectroscopic binaries (SB) with periods shorter than $\sim10$--$30$ 
{days}
and secondaries with masses in the range of $\sim0.01$--$0.1\, M_{\odot}$. 
We reconsider here their sample of binaries, focusing on 
116 
systems on the main sequence of the \gaia\ color-magnitude diagram, with mostly K-dwarf primaries. 
Using our recently devised algorithm 
to analyze the mass-ratio distribution of a sample of SBs we confirm the BDD existence and delineate its boundaries. 
For the K-dwarf APOGEE $1$--$25$ days
binaries, the companion-mass range of the BDD  is $\sim0.02$--$0.2\, M_{\odot}$.
The mass ratio distribution of the long-period (25--500 days) binaries does not show any dearth at the $q$-range studied. Instead, their distribution displays a linear increase in $\log q$, implying a tendency towards low-$q$ values.
The limits of the BDD do not coincide with the frequently used mass limits of the brown-dwarf population, sometimes defined as $0.013$ and $0.08\, M_{\odot}$, based on theoretically derived stellar minimum masses for burning deuterium and hydrogen in their cores.
Trying to draw the boundaries of the desert, we suggest either a wedged or trapezoidal shape.
We discuss briefly different scenarios that can account for the formation of the BDD, in terms of differentiating  between stellar secondaries and  planets in particular, and compare this desert to the Neptunian desert that can distinguish between Jovian planets and super Earths of short periods.
\end{abstract}

\section{Introduction}
\label{sec:intro}

Large surveys of stellar radial-velocities (RV), 
 performed in the quest for exo-planets
 \citep[e.g.,][]{queloz00,marcy00,pepe04,bouchy09},
have discovered  in the last few decades many spectroscopic binaries (SB) with large range of secondary masses. A few studies  noticed  a 
 dearth of companions  with $\sim0.01$--$0.1  \msun$ mass for systems with orbital periods shorter than $\sim100$ days
 \citep[e.g.,][]{marcy00,grether06}.
 \citet{ma14} assembled a literature-collected catalogue of 64 low-mass secondaries,  of 10--80 Jupiter masses 
($\mjup\sim10^{-3}\msun$), 
hosted by FGK-type  primaries.
 Their derived mass-period distribution suggested that at orbital periods shorter than 100 days,
secondaries with masses of 35--55 $\mjup$ are nearly depleted.
The observed dearth was named the Brown-Dwarf Desert (BDD), after the dim objects below $\sim0.08 \msun$, which cannot ignite hydrogen burning in their cores \citep[e.g.,][]{kumar62,kumar63,hayashi63}.
In this view, the few detected brown-dwarf (BD) {\it secondaries} with short orbital periods \citep[e.g.,][]{bouchy11,triaud17,grieves17,dossantos17,beatty18,hodzic18} 
are just oases found inside the `desert'. 

The BDD probably enables us to {\it statistically} distinguish between the populations of small stellar companions and massive planets. The very existence of the BDD might even indicate two different mechanisms of formation, below (planets) and above (stellar secondaries) the BDD 
\citep[e.g.,][]{grether06}. 

The observational characteristics of the BDD can shed some light on its origin \citep[e.g.,][]{marks17}. The shape and location of the desert in the period-secondary mass parameter space
\citep[e.g.,][]{schlaufman18}, 
and their dependence on the stellar mass and metallicity 
\citep[e.g.,][]{bouchy11,borgniet17,murphy18}, are of particular interest.
For example, it is hard to imagine that the BDD abruptly disappears for periods longer than some limiting period, as presented by some studies. Instead, we can expect a transition region at which the gap between the stellar secondaries and the massive planets gradually closes.

To further study the BDD and its borders, one needs an unbiased sample of SB systems obtained by a well-defined large survey with high enough precision, so observational effects can be estimated and corrected for. A large sample is not enough, since the unknown inclination of each binary does not allow deriving the secondary mass, even in cases where the primary mass can be estimated. As it is well known, assignment of some expected value of the inclination to all systems can distort the resulting mass-ratio distribution
\citep[e.g.,][]{mg92,heacox95}, 
and therefore in our case can twist the BDD boundaries.  

To overcome this problem one needs to apply statistical tools that utilize the assumed spherical symmetry of the binary inclinations
\citep[e.g.,][]{mazeh92, boffin93,cure15,swaelmen17}.
%
One of these tools is the modified mass  function
of single-lined spectroscopic  binaries
(SB1s), suggested recently by \citet{shahaf17}. They showed that this parameter, when derived for a sample of SB1s, follows the  underlying
mass-ratio distribution.  

A sample of SB1s that enabled the study of the BDD is the one recently released by APOGEE\footnote{APOGEE: http://www.sdss3.org/surveys/apogee.php}---the Apache Point Observatory Galactic Evolution Experiment
 \citep[][]{majewski17}.
APOGEE is an infrared spectroscopic survey of  Milky Way stars, with obtained spectra that cover the H-band wavelengths, from 1.51 to 1.69 $\mu$m, with a resolving power of $22,500$. The database contains spectra of over 146,000 stars, most of which were measured at several epochs. 
The binary sample became available with the thorough and careful analysis of \citet[][T16 henceforth]{troup16}. 

RVs were derived for most APOGEE spectra with a typical precision of 
$\sim100$ m/s
\citep[][]{nidever15}.
This precision enables the detection of stellar, sub-stellar and even giant planetary companions 
 \citep[e.g.,][]{price-whelan18, badenes18, el-badry18}. 
T16
performed an extensive study of APOGEE DR12, and focused on single-lined systems with at least 8 epochs taken during a period of $\sim3$ years. 

Their analysis yielded
 a  ``gold sample" of 
 382 SB1s, out of which  155  have been identified as having main-sequence (MS) primaries. 
T16 used their sample, 
considering binaries with MS and evolved primaries alike,
to notice that the BDD extends up to only 
orbital separations of 0.1--0.2 AU,
i.e., orbital periods of 10--30 days.

This work targets the BDD as manifested among the APOGEE systems with K-dwarf primaries, with the goal of delineating the BDD borders. In Section 2 we take advantage of the release of \gaia\ DR2 parallaxes and colors to  validate the classification of the primaries of the sample, and focus of a subsample of 
116
binaries with definite mass range, $0.5$--$1\, M_{\odot}$.
Although left with a relatively small sample, we 
try in Section~3
 to draw the shape and location of the BDD in the 
period--secondary-mass parameter plane. 
This is done by 
applying 
 the \citet{shahaf17} algorithm to derive the mass-ratio distribution of the sample in two period bins, in order  to determine the mass limits of the BDD, and 
by drawing either a wedged or trapezoidal shape for the boundaries of the desert.
Section~4 shortly discusses the meaning of the BDD in terms of planetary and binary formation.

%
%

\section{The Restricted APOGEE K-dwarf sample of Spectroscopic Binaries}

\begin{figure*}
	\centering
	\includegraphics[width=0.95\linewidth]{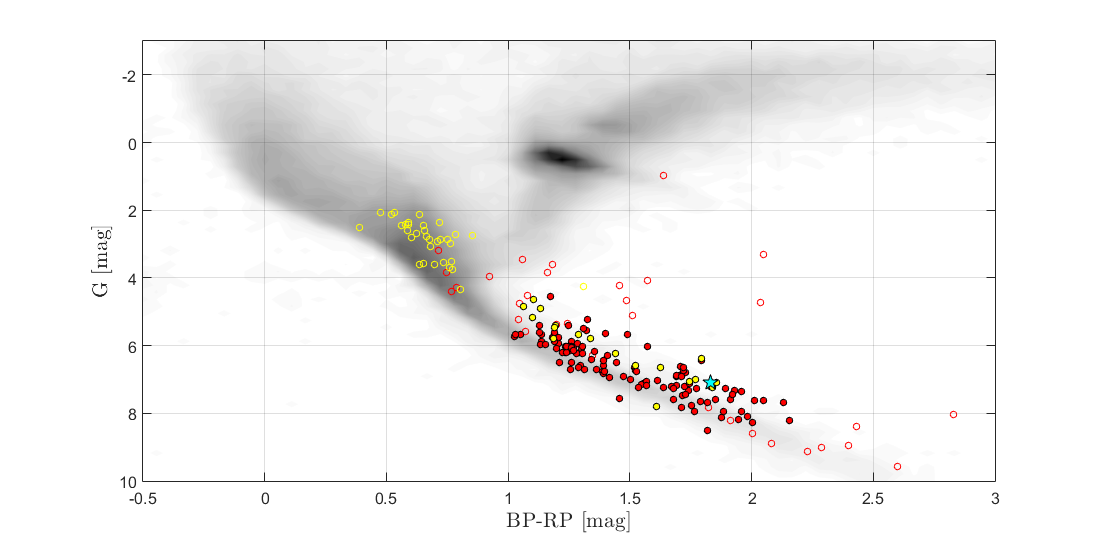}
	\caption{ 
Restricted APOGEE SB1 sample (filled red circles) on  \gaia\ CMD, overlaid on a gray-scale density map of \hipparcos\ stars, used as a proxy for the expected CMD in the solar neighborhood.
Empty red circles represent SB1s that were rejected from the analyzed sample in this work, due to their CMD position or mass estimates provided by Troup et al. (2016).
For completeness, 18 SB2s found by
\citet{el-badry18}  are presented as yellow circles.
Empty yellow circles represent SB2s that are not in the 
color-magnitude 
range of the restricted sample.
A cyan pentagram
marks the location of 2M13431527+1910491, a binary with minimal mass-ratio larger than 1 that was excluded from the restricted APOGEE K-dwarf sample.
}
\label{fig:cmdplot}
\end{figure*}

Figure~\ref{fig:cmdplot} displays a \gaia\ color-magnitude diagram 
\citep[CMD, see][]{lindegren18} of 
150 binaries
classified by T16 as having MS primaries, for which their parallaxes have been released by   \gaia\ DR2.
The gray-scale density map in the background represents the \hipparcos\ stars
\citep{lindegren97} measured by \gaia, used as a proxy for the expected CMD in the solar neighborhood. The MS and the giant branch are clearly separated in the diagram.

Figure~\ref{fig:cmdplot} suggests that a few of T16 MS stars are actually giants or subgiants. Those were removed from the sample we analyzed. 
Additionally, in order to have a small range of primary masses we limited the sample  to include  only systems redder than \gaia's BP$-$RP 1.0 mag, {\it and} required T16 mass estimates to be of $0.5$--$1\, \msun$. 
These requirements resulted in a subsample of 117 targets. 

One system, 2M13431527+1910491, had  an orbit of $\sim87$ days,
with a minimum mass ratio (see below) larger than unity. This system required further study, careful error estimation of the orbital parameters in particular, and therefore was excluded from the sample of this study. 
Its location on the \gaia\ CMD is marked by a cyan
 pentagram. 
We are left with 
 116 
stars that we define as the restricted sample, representing the K-dwarf SB1s detected by APOGEE. 

For completeness we added to Figure~\ref{fig:cmdplot} the double-lined spectroscopic systems (SB2) found in APOGEE spectra by \citet{el-badry18}. Out of 64 
SB2s with reported orbits,  51 were found in a cross-match with \gaia,  and  18 	
were located on the CMD near the SB1 restricted K-star sample.
As expected, most  of these SB2s are located slightly above the MS of the CMD.

%

\begin{figure}
	\centering
	\includegraphics[width=1 \linewidth]{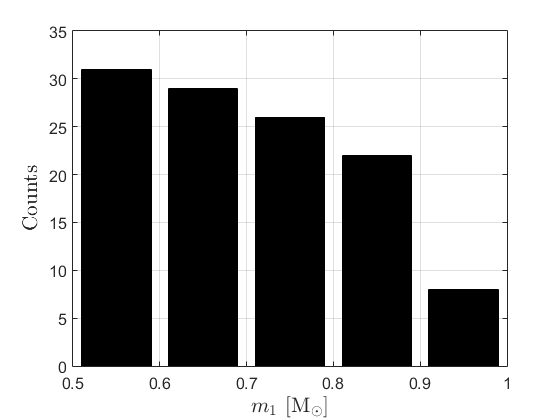}
	\caption{Primary-mass histogram of the restricted sample of 
		 116
		 K-type MS primaries. }
	\label{fig:massHist}
\end{figure}

A histogram of the primary masses, as reported by T16, of our restricted sample of SB1s, is shown in Figure~\ref{fig:massHist}. 

%

\begin{figure*}
	\centering
	\includegraphics[width=0.75 \linewidth,trim={10 0 0 10 },clip]{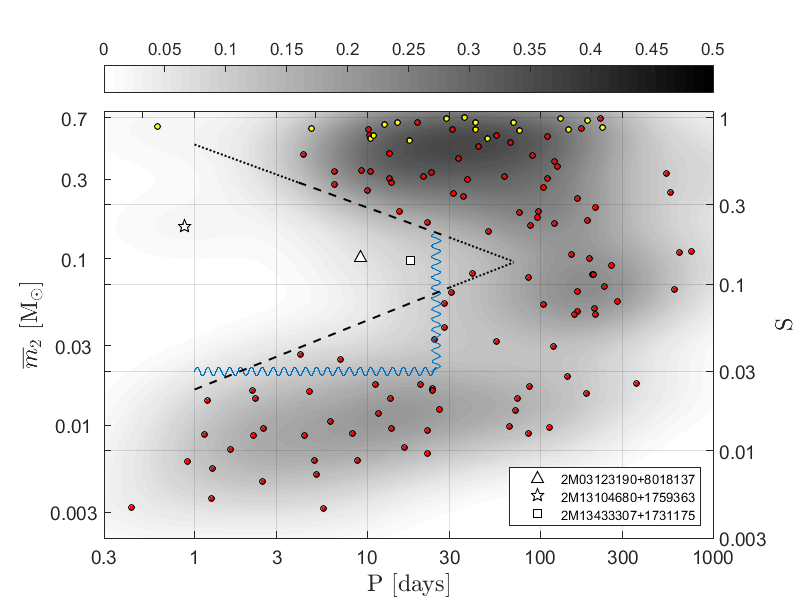}
\caption{ Estimated secondary mass, $\overline{m}_2$, vs.~binary period, $P$, plotted on a logarithmic scale. The horizontal axis on the right shows the value of the modified mass function, $S$. A kernel density estimate of the points appears in gray on the background.
The restricted sample binaries are marked by red points, except three outliers that appear in white (see legend). Additionally, yellow points mark the 18 SB2s obtained from \citet{el-badry18}, that were found to be located  close to the restricted sample on the \gaia\ CMD.
The figure suggests two alternative shapes of the desert borders (see text): a wedged-shape, with fitted slopes (dashed and dotted lines), and an eye-drawn trapezoid 
with blue serpentine line).
}
\label{fig:SvsP}
\end{figure*}

\section{The brown-dwarf desert and its boundaries}
 
For each of the  116 binaries
of the restricted sample, we use its primary mass, $m_1$, and the mass function, $f(m_1)$, reported by \citet{troup16}, to calculate the reduced mass function, $y$, which can be expressed as: 
%
\begin{equation} \label{EQ: norm mass function}
y \equiv \frac{f(m_1)}{m_1} = \frac{q^3}{(1+q)^2} \,  sin^3i \, ,
\end{equation}
%
where $q$ is the mass ratio and $i$ is the orbital inclination angle.

Since the reduced mass function, $y$, is a combination of two unknowns, $q$ and $i$, one cannot directly infer the mass ratio of each system.
However, 
as is well known, each value of $y$ defines a minimal possible mass ratio,  $\qmin$, that can be determined by setting the inclination angle $i$ in equation~(\ref{EQ: norm mass function}) to be $90^{\circ}$.

To facilitate the derivation of the mass-ratio distribution \citet{shahaf17} introduced a new observable, $S$, coined the `{\it modified} mass function',  and showed that the derived $S$ distribution is similar to the underlying mass-ratio distribution of the sample. We therefore derived
$S$ for each target in the restricted sample, 

\begin{equation} \label{EQ:S}
S = 1 - \int_{\qmin}^{1}{\sqrt{1 - y^{2/3} \,{(1+q)^{4/3}}{q^{-2}}}\,dq} \, ,
\end{equation}
and used its observed distribution as a proxy for  the underlying mass ratio (see below).

Finally, we plot in Figure~\ref{fig:SvsP} the results on a 
period--secondary-mass
diagram, in order to obtain a clear view of the BDD. 
As a proxy for the secondary mass we use $\overline{m}_2 $, 
\begin{equation}
\overline{m}_2\equiv  0.7 \cdot S \,\,\,\small{\msun} \, , 
\end{equation}
namely $S$ multiplied by the typical primary mass value of the sample. 
For the 18 SB2 from \citet{el-badry18} we used the derived mass ratio instead of the modified mass functions. 
Kernel density estimation of the sample \cite[see][]{botev10}  appears in gray-scale as a background.

\subsection{The Mass-Ratio Distribution of the K-dwarf SB1s}

We used the restricted sample to derive the mass-ratio distribution of two separate period bins, of  
1--25 and 25--500 days,
which resulted in sub-samples of 49 and 57 binaries, respectively. 
Each period bin was analyzed separately, by approximating the probability density function with a set of logarithmically equally-spaced boxcar functions.
According to the Rice rule \cite[]{terrell85}, and since the sample sizes are of $\sim50$ binaries, we use eight bins that span $\log q$ from $-2.4$ to $0$, at spacing of $0.3$.  

The mass-ratio distribution was then fitted by maximizing the likelihood of the sample, using an ensemble MCMC method \cite[\texttt{emcee},][]{foreman-Mackey2013}.
We assumed that the threshold for T16 APOGEE detection was an RV semi amplitude of 200 m/s, and corrected the induced observational bias accordingly
 \citep[see][]{shahaf17}. 

\begin{figure*}
	\centering
	{\includegraphics[width=0.9\linewidth]{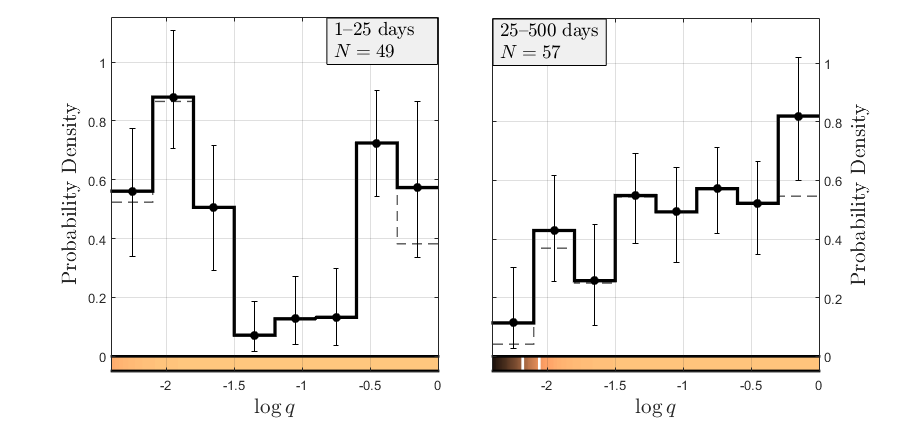}} 
\caption{Derived mass-ratio distribution in two selected period bins. The thick black line represents the fitted box-car shaped model. 
The fitted bins are of constant width of $0.3$ and span the range
from $-2.4$ to $0$ in $\log q$.
The error bars represent $1\sigma$ confidence interval. 
The gray dashed line shows the fitted distribution before correcting for detection bias. The expected fraction of detected binaries as a function of $\log{q}$, given the assumed threshold of 200 m/s, appears as a bar below the fitted model, where black is no detection and copper is 100\% detection. 
The two white lines on the bottom of the right panel represent the 50\% and 75\% percent detection probability. The sample presented on the left panel is of SB1s with shorter period range, and therefore its detection probability $\gtrsim 90\%$ throughout the fitted range.
}
\label{fig:mrdfig}
\end{figure*}

The resulting mass-ratio distributions are plotted in Figure~\ref{fig:mrdfig}. The fitted normalized distribution appears as a dashed gray line. 

The thick black line represents the distribution corrected for systems undetected due to small RV amplitudes \citep[see][]{shahaf17}. The magnitude of the correction is shown as a colored line below each figure. 

The distribution derived in Figure~\ref{fig:mrdfig} was based on the T16 SB1 sample only, ignoring the SB2s found by \citet{el-badry18}. As all the SB2s had large mass ratio, close to unity, the resulting $q$-distribution was strongly biased at its high end. 
To correct for this effect we applied a 50\% correction to the largest $q$ bin. Still, this correction is highly arbitrary and therefore the values of last bin in each of the derived distributions are quite uncertain, and should not be used in any astrophysical discussion before further study.

The left panel, with the mass-ratio distribution of the short-period systems, clearly displays the BDD, extending over three bins at $-1.5 \lesssim \log q \lesssim -0.6$. In terms of the typical secondary mass, the desert is at
%
\begin{equation}
0.02 \lesssim m_2/\msun \lesssim 0.2 \, .
\end{equation}
%

The mass ratio distribution of the long-period  (25--500 days)
binaries does not display any dearth at the $q$-range studied. Instead, the resulting distribution suggests a linear increase in $\log q$, implying 
a tendency towards low-$q$ values.
%

\begin{figure*}
	\centering
	{\includegraphics[width=0.9\linewidth]{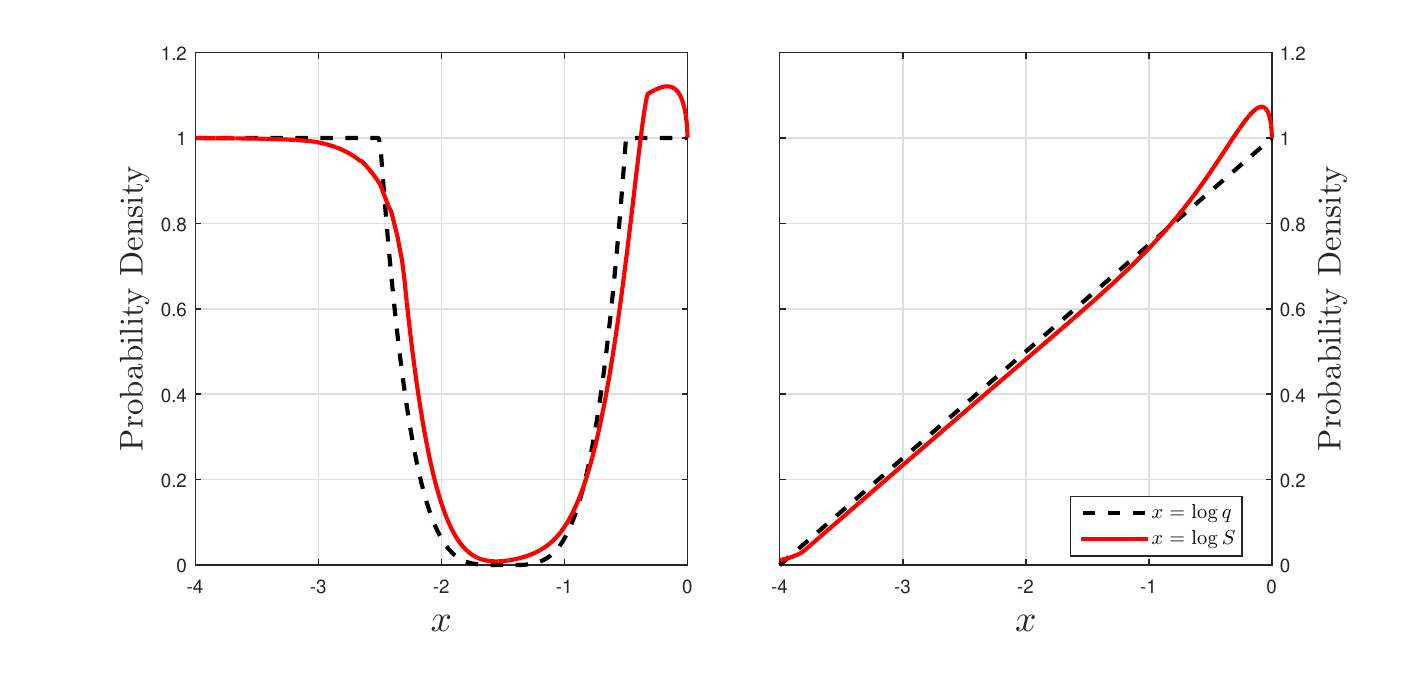}} 
\caption{Desert-type (left) and linear in log-$q$ (right) mass-ratio distributions (black dashed) compared with the derived distribution of the modified mass function (red).
}
\label{fig:simulation}
\end{figure*}

Figure~\ref{fig:simulation} 
compares the modified mass function distribution \citep{shahaf17} both with a mass-ratio distribution of a desert-shape and a linear distribution in $\log q$.
%
%
The figure demonstrates the capacity of the modified mass function
to follow the mass-ratio distribution in both cases. 
Note the small excess of the $ S$  distribution at  the edge of the studied range, which is inherent feature of the modified mass function
(see equation~2).
A code to derive the modified mass function is available on-line.\footnote{\url{https://github.com/saharsh1/BinaryMassFunction}}

\subsection{The boundaries of the BDD}

In this subsection we discuss two alternative ways to draw the boundaries of the BDD. Unfortunately, the SB1 sample is not large enough to differentiate between the two shapes. The two resulting shapes are  similar, and the difference does not have an obvious impact on the discussion of the astrophysical implication of the desert. 

\subsubsection{Wedged-Shape Desert}
Following \citet{mazeh16}, we tried to delineate the low- and high-mass BDD borders by fitting the observed occurrence rate with a sigmoid function, relative to a linear borderline 
\begin{equation}
\mathcal{M}_2 
= a \mathcal{P}
+ b  ,
\end{equation}
where $\mathcal{M}_2$ and $\mathcal{P}$ represent $\log \frac{\overline{m}_2}{\scriptsize{\msun}}$ and $\log \frac{P}{\small{\text{day}}}$, respectively.

We  modeled  the probability for an observed system with measured 
$\mathcal{M}_2$ and $\mathcal{P}$ to be located at a distance $d$  from the line as 
%
\begin{equation}
\mathcal{F}\bigg( \mathcal{M}_2,\mathcal{P}\,\, ;\,\,a,b\bigg) = A \bigg\{ \bigg[ 1+e^{-{d}/\delta} \bigg]^{-1} + \Delta \bigg\} \ ,
\label{EQ: sigmoid model}
\end{equation}
%
where  $\delta$ is the transition-steepness, $\Delta$ the density inside the desert and $A$ is a normalization factor. The distance $d$ is taken as the
distance of $\big(\mathcal{P},\mathcal{M}_2\big)$ from the borderline, so that $d>0$ outside the desert. 

The parameters $a$, $b$, $\delta$ and $\Delta$ were estimated by maximizing the likelihood of the model for the location of the upper and lower borders, given the sample, using an MCMC ensemble sampler \cite[see][]{foreman-Mackey2013}.

The lower border was fitted with the 34 systems that have periods in the range of 1--30 days
and $S<0.1$. The upper border was fitted with the 15 systems that have periods in the range of 4--30 days
and $S>0.15$. 
The derived borders are represented in Figure~\ref{fig:SvsP} by a dashed line. The dotted line shows an extrapolation of the borders, such that the borderline covers the 1--70 day range.

\begin{table}
	\centering
\begin{tabular}{l c c c c }
	\hline \hline
	 & a & b & $\delta$ & $\Delta$ \\ 
	\hline \hline
	Upper &  $-0.41^{+0.20}_{-0.25}$& $-0.28^{+0.26}_{-0.23}$ & $0.005^{+0.013}_{-0.004}$  & $<0.07$ \\ 
	\hline 
	Lower &  $0.40^{+0.12}_{-0.11}$ & $-1.80^{+0.11}_{-0.09}$ & $0.0025^{+0.0042}_{-0.0021}$ & $<0.002$ \\ 
	\hline \hline
\end{tabular} 
\caption{Fitted borderline parameters. The fitted probability density inside the desert, $\Delta$, was consistent with 0, therefore it's $3\sigma$ upper limit is presented. }
\end{table}
 
\subsubsection{Trapezoid-Shape Desert}
As the lower boundary of the BDD  is constrained by a small number of systems, we offer in Figure~\ref{fig:SvsP} alternative borderlines---horizontal and vertical straight lines. Their locations were chosen, somewhat arbitrarily at $S=0.03$ and $P=25$ days.
The upper boundary is left as before. The trapezoid-shape limits are drawn with blue serpentine line.


\subsection{Three Binaries inside the BDD}
%
Figure~\ref{fig:SvsP} shows  three binaries residing deep within the proposed BDD boundaries.
Their fitted orbital parameters, as of all other ``gold sample" binaries, were made available on-line by T16. For completeness we briefly summarize some of their properties in Table~\ref{table:BDDoutliers}.  
To validate the results of T16, we re-fitted the orbital solution of the three systems and searched for center-of-mass acceleration, which may indicate a gravitational pull by a distant stellar companion. The table gives our $P$, $K$ and $e$ determination, all of which consistent with T16 (see below). 
We found that only the RVs of 2M13433307+1731175 displayed  a possible acceleration of $-0.053\pm0.011$ km/s/day.

The  orbital solution of T16  for 2M13104680+1759363 displayed a large uncertainty, of over $800$ km/s, on the RV semi-amplitude.  Our analysis yielded an orbital solution consistent with that of T16 but with a substantial smaller uncertainty, given in Table~\ref{table:BDDoutliers}. Since there were only 9 available RVs, we did not fit an acceleration component to this target.

\begin{table}
\begin{tabular}{r r r r r}
	\hline \hline
   {APOGEE ID}	& {N} &	{P [day]}    & {K [km/s]}      & \thead{$e$} \\ 
	\hline 
	\hline
	2M03123190& 30 & $9.1756$     & $8.10$   &   $0.414$  \\ 
	+8018137  &    & $\pm0.0033$  & $\pm0.12$ &  $\pm0.011$  \\ 
	\hline 
	2M13104680& 9&$0.878933$ & $24.0$ &  $<0.03$  \\ 
	+1759363& &$\pm0.000094$  &$\pm1.0$  &    \\ 
	\hline 
	2M13433307& 13&$17.76$  & $6.31$ &  $0.549$  \\ 
	+1731175& &$\pm0.15$ & $\pm0.25$ &  $\pm0.030$ \\ 
	\hline \hline
\end{tabular} 
\caption{Number of measurements, orbital period, RV semi amplitude and eccentricity of the three binaries located inside the proposed BDD borders.}
\label{table:BDDoutliers}
\end{table}

\section{discussion}

We have presented here a detailed analysis of the BDD, based on
a restricted sample of 
116
APOGEE K-dwarf SB1s detected by  T16. 
For binaries with $\sim$1--25 day period, the mass-ratio distribution reveals a dearth of secondaries with mass of $\sim 0.02$--$0.2\, M_{\odot}$.  For binaries with longer periods 
(25--500 days)
no desert is seen, and the derived mass-ratio distribution of the sample
 tends towards low-$q$ values.
The period limit of the BDD, $\sim25$ days,
corresponds to orbital separation of $\sim0.15$ AU, as pointed out already by T16.
In the \MP\ plane, the desert is probably of a wedged or trapezoidal shape.

The mass limits of the BDD do not coincide with the generally accepted mass limits of the BD population, defined as $0.013$ and $0.08\, M_{\odot}$, based on the minimum stellar mass required for deuterium  and  hydrogen burning in their cores, respectively
\citep[e.g.,][]{burrows01,auddy16}, but see also \citet{forbes19}. 
In particular, the findings are in contrast with the claim that the mass upper limit of the BDD coincides with the
stellar/BD mass transition. This claim was difficult to explain, 
since it is not clear how the nuclear astrophysics of the stellar core
can also surface in the binary formation of short-period systems.


One may think of two types of interpretation of the BDD:
\begin{enumerate}
\item
As an outcome of binary/planetary formation, the primordial density of binaries in the \MP\ plane was flat, with no BDD.
 The desert was formed by some later mechanism that caused objects found in the BDD to
`move away' from the `restricted area'
\citep[e.g.,][]{armitage02,damiani16,grunblatt18,vick19}. 
%
This could happen by

\begin{itemize}
\item
Enlarging the mass of the companions found in the BDD, effectively pushing the systems up in the \MP\ plane.
\item
Stripping part of the companions' mass, effectively pushing systems down in the \MP\ plane.
\item
Spiraling the secondaries into the primary, maybe by tidal interaction.
\item
Pushing the secondaries out towards larger orbits, maybe by tidal interaction.
\end{itemize}

\item
The BDD is an outcome of a gap between two different formation mechanisms---binary formation with stellar-mass secondaries, and planetary formation with Jovian and smaller masses
\citep[e.g.,][]{ma14,chabrier14}.
The binary formation has a lower mass limit larger than the upper mass limits of planets. 
\end{enumerate}

Note that if we accept the first class of interpretations, which assumes that the objects at the BDD were pushed out of the restricted area, we can naturally expect a wedged or trapezoidal shape. The effectiveness of the mechanism that clears up the BDD area, whatever its origin might be, could get weaker for longer periods and larger orbital separation. On the other hand, if we adopt the second interpretation, we still need to explain why the slope of the upper boundary, which is the lower envelop of the stellar secondaries, is clearly negative, whereas the slope of the lower boundary, which is the upper envelop of the planets, is definitely not negative. 

The (almost) generally accepted paradigm is that planetary and binary formations operate with two different mechanisms. Binary formation is
probably driven by fragmentation of the early contracting protostar
\citep[e.g.,][]{bate97,bonnel06, bonnel08,riaz18}, 
whereas planets are formed by coagulation of small planetesimals in an accretion disc around the star \citep[e.g.,][]{pollack96,goldreich04,levinson10}. 
A natural consequence of the duality of the binary and planetary formations is the second interpretation of the BDD. Accordingly, the desert is the product of a lucky coincidence. For {\it short-period} orbits, the upper mass limit of planetary formation is smaller than the lower mass limit of stellar formation. This enables us to distinguish between the short-period planets and stellar secondaries. 

Provided this is true, our analysis indicates that in the context of the BDD the distinction between a BD and a planet at $13\, M_{\rm Jup}$, based on nuclear ignition of deuterium, is not very useful. We have shown that for the APOGEE K-dwarf sample, the BDD extends down only to about 
$\sim 20\, M_{\rm Jup}$. 
In any case, `planet' should be an attribute associated with a small-mass object orbiting a star, while a BD is termed to describe an object that fails to ignite hydrogen in its core, regardless of its dynamical properties. 

In the analysis presented above three binaries clearly stand out in the middle of the desert. Their isolated locations deserve a special attention, based on the assumed mechanism behind the BDD. Were those systems pushed into the desert after their formation, maybe by some interaction with a third star through the Kozai-Lidov mechanism? This idea was suggest by \citet{fabrycky07}
to account for the  formation of {\it all} short-period binaries \citep[see also][]{mazeh79}.
If so, we should find some evidence for those oases to have third distant faint companions, 
as in the case of HD 41004 Bb \cite[see][]{zucker03}. 

Alternatively, we should find that the frequency of binaries inside the BDD with third companions is higher than it is for systems outside the desert.
Very recent study of \citet{fontanive19} suggests that this is the case for close giant planets, not necessarily inside the BDD.
Currently, there are $\sim120$ known planets that orbit a star that has a distant  stellar companion, listed in a catalogue\footnote{\url{https://www.univie.ac.at/adg/schwarz/multiple.html}} by \citet[][]{schwarz16},
but a careful analysis of the frequency of such systems is still not available. 

Conversely, the desert might be  not so dry at the first place
\citep[see][and references therein]{carmichael19}. Some binaries could be formed at the middle of the BDD, because of some scarce initial conditions. 
%
%
We need more similar systems, followed by in-depth studies, to understand the nature of these special cases
\citep[e.g., WASP-128b,][]{hodzic18}.

The existence and characteristics of the BDD should be compared with the Neptunian desert found around G-dwarf primaries \citep[e.g.,][]{ Szabo11, mazeh16}, based mainly on the {\it Kepler} space-mission discoveries.
%
%
The Neptunian desert 
also has a wedged shape at the \MP\ plane, centered around $\sim 0.1\, M_{\rm Jup}$, whereas the BDD of the K-dwarf APOGEE is {\it centered} around $\sim70\, M_{\rm Jup}$. 

Similar to the possible role of the BDD, the Neptunian desert can distinguish between Jovian planets and super Earths of short periods. 
The formation of the desert is still debated
\citep[see, for example,][for theoretical discussion]{arieh16,ionov18,owen18}. 
The recent study of \citet{szabo19} that considered the boundaries of the Neptunian desert and their dependence of the primary mass and temperature should shed some light on its origin. 
Adopting the right 
interpretation, like in the BDD case, depends on the formation scenarios of the populations on the two sides of the desert.

Do the two deserts, separated by almost three orders of magnitude, send us similar messages, namely that they both were formed by mass limits of two different mechanisms? If so, then we are now facing three mechanisms of formation---stellar secondaries, Jovian {\it and} super-Earth planets. Again, in the short-period domain, the two deserts help us distinguish between the three populations, and should be used as clue for detailed modeling of their formation.

A critical test of this interpretation of each of the two deserts is the dependence of their location and shape on stellar and environment features, stellar mass and metallicity in particular
\citep[e.g.,][]{bouchy11,guillot14}. 

Large samples of short-period binaries will be available with the releases of the \gaia\ SB1 systems in the next few
years.\footnote{https://www.cosmos.esa.int/web/gaia/release}
Large samples of transiting planets and BD secondaries around different types of primary stars are being discovered by TESS\footnote{TESS: https://tess.gsfc.nasa.gov/} \citep[][]{ricker14}, 
such as HD 202772A b \citep{wang19} and HATS-71b \citep{bakos18}.	
The new findings will enable us to better understand the nature of both deserts in a few years. 

\section*{ACKNOWLEDGEMENTS}
 We are indebted to the APOGEE team and to Troup and his group for the fantastic set of spectra and the careful analysis of the spectroscopic binaries. This study would not be possible without their seminal work.
Special thanks to Simchon Faiglar for careful reading of the manuscript and enlightening comments. This research was supported by Grant No. 2016069 of the United States-Israel Binational Science Foundation (BSF) and by the Israeli Centers for Research Excellence (I-CORE, grant No. 1829/12).
  %
%
\bibliographystyle{mnras}
\bibliography{MR_bib}

%

\bsp	
\label{lastpage}
\end{document}